# Factors limiting ferroelectric field-effect doping in complex-oxide heterostructures

L. Bégon-Lours[1,♦], V. Rouco[1], Qiao Qiao[2], A. Sander[1], M. A. Roldán[3], R. Bernard[1,♣], J. Trastoy[1], A. Crassous[1], E. Jacquet[1], K. Bouzehouane[1], M. Bibes[1], J. Santamaría[1,3], A. Barthélémy[1], M. Varela[3] and Javier E. Villegas[1,*]

[1]Unité Mixte de Physique, CNRS, Thales, Univ. Paris-Sud, Université Paris Saclay, 91767 Palaiseau, France
[2]Oak Ridge National Laboratories, Oak Ridge TN 37840, USA
[3]GFMC, Dpto. Física de Materiales, Universidad Complutense de Madrid, 28040, Spain

* corresponding author: javier.villegas@cnrs-thales.fr
♦ present address: MESA+ Institute, Inorganic Materials Science, Enschede, The Netherlands
♣ present address: Fonctions Optiques pour les Technologies de l'Information, UMR 6082, CNRS, INSA de Rennes, Rennes, France

Ferroelectric field-effect doping has emerged as a powerful approach to manipulate the ground state of correlated oxides, opening the door to a new class of field-effect devices. However, this potential is not fully exploited so far, since the size of the field-effect doping is generally much smaller than expected. Here we study the limiting factors through magneto-transport, scanning transmission electron and piezo-response force microscopy in ferroelectric/superconductor (YBa$_2$Cu$_3$O$_{7-\delta}$ /BiFeO$_3$) heterostructures, a model system showing very strong field-effects. Still, we find that they are limited in the first place by an incomplete ferroelectric switching. This can be explained by the existence of a preferential polarization direction set by the atomic terminations at the interface. More importantly, we also find that the field-effect carrier doping is accompanied by a strong modulation of the carrier mobility. Besides making quantification of field-effects via Hall measurements not straightforward, this finding suggests that ferroelectric poling produces structural changes (e.g. charged defects or structural distortions) in the correlated oxide channel. Those findings have important consequences for the understanding of ferroelectric field-effects and for the strategies to further enhance them.

# I. INTRODUCTION

Ferroelectric field effects [1–4] can be achieved in heterostructures in which a ferroelectric gate and a strongly correlated oxide channel are combined (e.g. a superconducting cuprate [2,5], a half-metallic manganite [6] or a Mott insulator [7]), following the scheme of a transistor. As the bound charges in the ferroelectric are screened by free-charges in the channel, carriers are either accumulated or depleted depending on the direction of the ferroelectric polarization. Owing to the extreme sensitivity of correlated oxides to carrier density variations [8], they can be driven across boundaries in their phase diagram upon ferroelectric switching by a gate voltage [4,9]. This effect ultimately enables electrical control of functionalities [5,10–12]. In this context, the advantages of ferroelectric versus dielectric or liquid-electrolyte gating [13–16] include the nonvolatile character of the doping [2,17], the possibility to produce it selectively over nanometric lateral scales [5,10,11], and the short switching times involved (below ~ns). These characteristics make ferroelectric gates the most promising for new application concepts [8].

However, ferroelectric field effects are generally weaker than those found in liquid-electrolyte gating [13–15]. In principle, the field-effect doping size should correlate with the magnitude of the bulk ferroelectric polarization $P$, which amounts nearly a hundred $\mu C \cdot cm^{-2}$ in oxide ferroelectrics, for example in $BiFeO_3$ (BFO). In particular, a sheet carrier density modulation $\Delta n_{2D}$ up to $2P/e$ ~ $10^{14}$-$10^{15}$ cm$^{-2}$ is expected upon full ferroelectric switching [9], allowing for significant modulation of parameters such as the critical temperature $T_C$ in superconductors [2,5,10] or the saturation magnetization in ferromagnets [18]. However, quantitative measurements show that the induced $\Delta n_{2D}$ is only a small percentage of $2P/e$ [2,5,7,10]. This fact suggests that a number of complex factors and mechanisms are simultaneously at play, and that the potential of ferroelectric field-effects is far from being fully exploited.

Here we study a model system: superconducting ($YBa_2Cu_3O_{7-\delta}$)/ferroelectric ($BiFeO_3$) heterostructures in which we observe a large ferroelectric modulation of the superconducting critical temperature accompanied by a consistently large modulation of the carrier density. The doping efficiency $e\Delta n_{2D}/2P$ reaches up to ~20% which, despite being among the highest reported in the literature, is far from full efficiency. Furthermore, we find that $e\Delta n_{2D}/2P$ unexpectedly decays with increasing $YBa_2Cu_3O_{7-\delta}$ thickness.

Using ferroelectric characterization at the nanoscale, we rule out that poor ferroelectric properties near the interface may account for any strong reduction of the doping efficiency. However, we find that that full ferroelectric switching is not achieved upon ferroelectric poling. This can be explained by the observation of a preferential polarization direction, which originates at pinned dipoles set by the atomic terminations at the interface. In addition, and very importantly, the analysis of the transport experiments shows that the carrier density modulation is accompanied by a substantial modulation of the carrier mobility − an effect generally overlooked, but whose relevance has been brought to light by recent studies on ferroelectric/rare-earth nickelate heterostructures [19]. We show here that if the carrier mobility variations are neglected in the Hall measurements analysis, the ferroelectric doping efficiency can be severely misestimated, particularly when the channel is thicker than the Thomas-Fermi screening length. This is indeed at the origin of the unexpected doping efficiency decay inferred from a plain analysis of the Hall coefficients. Chiefly, the strong mobility modulation suggests that the carrier modulation is accompanied by structural changes in the oxide channel, which could be for example structural distortions [19] or likely disorder in the form of charged point defects (such as oxygen vacancies) that would contribute to the screening of the electric field.

The ingredients invoked above (ferroelectric asymmetry controlled by the terminations at polar interfaces [20], screening by oxygen vacancies [21]), as well as the experimental conditions (correlated oxide channel thicker than the Thomas Fermi screening length) are ubiquitous in ferroelectric/correlated oxide interfaces and their studies, and thus the impact of the discussed factors on ferroelectric field-effects doping and their quantification should be quite general.

## II. EXPERIMENTAL METHODS.

### A. Sample growth

Epitaxial heterostructures with a fixed Mn-doped BiFeO$_3$ (BFO) thickness (30 nm) and varying YBa$_2$Cu$_3$O$_{7-\delta}$ (YBCO) one (2 to 6 unit cells) with a few (typically 4) unit-cells buffer layer of PrBa$_2$Cu$_3$O$_7$ (PBCO) were grown on (100) SrTiO$_3$ (STO) substrates by pulsed laser deposition, using a KrF 248 nm excimer laser with an energy density of ~3.7 J/cm$^2$, and a repetition rate of 1 Hz. PBCO is a large-gap semiconductor isostructural to YBCO. The presence of this buffer layer enhances the superconducting properties of the heterostructures [17]. The *in situ* growth of the different materials was done at 700 ºC for PBCO and YBCO, and subsequently at 560 ºC for BFO, in a 0.36 mbar pure oxygen atmosphere. After BFO growth, the samples were cooled-down to room temperature in 800 mbar of pure oxygen. Under the used conditions, the epitaxial YBCO and PBCO layers are *c*-axis (001) oriented. BFO grows under compressive strain on the YBCO/PBCO//STO stack, which as we showed earlier [17] induces in BFO a monoclinic structure with c-axis lattice parameter $c = 4.07$ Å instead of $a = b = c = 3.96$ Å observed for bulk BFO. Further details on the growth conditions and structural characterization can be found elsewhere [5,17].

### B. Electrical transport

Multi-probe transport bridges were defined using a combination of standard optical lithography and Ar$^+$ ion etching. The bridges width is 15 or 20 μm, depending on the sample, and the distance (longitudinal) between voltage probes is 40 μm. Samples were wedge-bonded using Al wire. A four-probe configuration was used to obtain the longitudinal R$_X$ and transverse R$_Y$ (Hall) resistances from the voltage (respectively V$_X$ or V$_Y$) measured under an injected current I$_X$. Measurements were carried out either in a close-cycle refrigerator equipped with a

~1 T electromagnet or in a liquid He cryostat equipped with a ~0.6 T electromagnet, the magnetic field being perpendicular to the films plane.

### C. Electron Microscopy

Scanning transmission electron microscopy (STEM) combined with electron energy-loss spectroscopy (EELS) studies were carried out in a Nion UntraSTEM200 operated at 200 kV and equipped with a fifth order aberration corrector and a Gatan Enfinium EEL spectrometer, at Oak Ridge National Laboratory. Atomic resolution EELS maps were produced by performing multiple linear least square (MLLS) fits on unfiltered, raw EEL spectrum images. Specimens were prepared by conventional methods, including mechanical polishing and Ar ion milling.

### D. Piezo-response microscopy

Ferroelectric domains were "written" by applying $V_{DC}$ between the scanning tip of a conductive-tip atomic force microscope (CT-AFM) and the YBCO layer, as discussed elsewhere [5]. The tip was electrically grounded. The remnant polarization state was "read" by measuring the piezo-response under an AC voltage excitation of amplitude 1 $V_{PP}$ at a frequency of 35 kHz. Subsequent DC poling and AC reading can be achieved with the tip fixed a particular spot, to obtain cycles as those shown in Fig. 4 (c). For these, $V_{DC}$ is set to the target value and then to 0V at a rate ~1 V s$^{-1}$. The piezo-response is measured in the remnant state with an AC voltage excitation of amplitude 1 $V_{PP}$.

## III. EXPERIMENTAL RESULTS

### A. Transport experiments

A multi-probe bridge was lithographed on the samples [see sketch, inset of Fig. 1 (a)] to study the ferroelectric field-effect doping through resistance and Hall effect measurements in the virgin state and at remanence after poling the BFO via CT-AFM. Because YBCO is a hole-doped superconductor, carrier depletion or accumulation is expected depending on whether the net polarization points towards ($P_\downarrow$) or away ($P_\uparrow$) from the interface, respectively.

The main panel of Fig. 1 (a) shows resistance *vs.* temperature for different states of the same sample. In the depleted state ($P_\downarrow$, blue curve), $T_C$ is only slightly lower than in the virgin state (green curve). This suggests that in the virgin state the ferroelectric polarization is similar to that set after poling the BFO towards the interface ($P_\downarrow$, hole depleted state). In contrast, $T_C$ dramatically increases in the hole accumulated state ($P_\uparrow$, red curve), as expected [2,5]. R(T) and Hall coefficient $R_H$ measurements at T>$T_C$ (either at 80 K or 100 K for different sample series) were carried in the virgin/doped/depleted states for series of samples with varying YBCO thickness, to obtain $T_C$ *vs.* $1/eR_H$ (*e* is the electron charge) shown in Fig. 1 (b). In YBCO, the superconducting critical temperature $T_C$ depends on the carrier concentration *n* [22–25]. The expected $T_C(n)$ for bulk YBCO is plotted in Fig. 1 (b) (dashed line). The experimental T$_C$ *vs.* $1/eR_H$ roughly follows $T_C(n)$. The most important observation is that, for every sample (that is, for each symbol shape), the critical temperature variation $\Delta T_C$ between different states (symbol colors) correlates with the Hall coefficient variation $\Delta(1/eR_H)$ following the $T_C(n)$ trend. This shows that, as discussed earlier [2,5], $T_C$ variations are associated with carrier density variations $\Delta n$ produced upon ferroelectric switching.

Fig. 1 (c) displays the product of the YBCO thickness ($t_{YBCO}$) with the variation $|\Delta(1/R_H)|$ produced upon switching from the depleted to the doped state (blue-red symbols), as a function of $t_{YBCO}$. If we assume − following the observations on Fig. 1(b) − that $1/eR_H$ equals the average carrier density in the YBCO layer, then $t_{YBCO} \cdot |\Delta(1/R_H)| = \Delta n_{2D} \cdot e$, with $\Delta n_{2D}$ the YCBO *sheet* carrier density variation. In principle, $\Delta n_{2D} \cdot e$ must equal the polarization variation produced by ferroelectric switching, that is, $\Delta n_{2D} \cdot e = \Delta P = P_\uparrow - P_\downarrow$. While the maximum $\Delta n_{2D} \cdot e$ ~30 μC cm$^{-2}$ we observe is large as compared to most data in the literature [2,7], it is only ~20% of $2P_z^{BFO}$ ~ 120 μC cm$^{-2}$ expected for a full switching of the bulk BFO polarization along the [001] direction. Furthermore, and strikingly, $t_{YBCO} \cdot |\Delta(1/R_H)|$ steadily decreases with increasing $t_{YBCO}$. This is completely unexpected. The reason is that $\Delta P = P_\uparrow - P_\downarrow$ must not depend on the YBCO thickness since, as discussed below, the structural and ferroelectric properties of BFO are similar for 3 u.c.<$t_{YBCO}$< 6 u.c. Therefore, $t_{YBCO} \cdot |\Delta(1/R_H)|$ should be constant. As we explain below, the unexpected thickness dependence observed in Fig. 1 (c) indicates that the measured $t_{YBCO} \cdot |\Delta(1/R_H)|$ does not exactly reflect the sheet charge carrier density variation.

**B. Electron Microscopy**

We conducted electron microscopy experiments in the virgin (as-grown) samples to investigate the link between doping efficiency, structural and ferroelectric properties at the nanoscale.

Fig. 2 and Fig. 3 show a summary of electron microscopy results representative of a series of BFO/YBCO/PBCO/STO heterostructures with YBCO thicknesses varying between 3 and 6 unit cells. Atomic resolution STEM-EELS confirm the high quality local crystalline structure of the samples. Lower magnification images (not shown) demonstrate flat and continuous layers over micrometric lateral distances. Fig. 2(a) displays a cross section high angle annular dark field (HAADF) image across the BFO/YBCO/PBCO/STO stacking from a BFO-Mn (30 nm)/YBCO(3 u.c.)/PBCO(4 u.c.)/STO sample. Interfaces are coherent and no major defects such as interface dislocations or secondary phases are observed. EELS imaging was used to ascertain the chemical integrity of the layers. Fig. 2(b) shows a series of atomic resolution chemical maps obtained from EEL spectrum images including the Ti $L_{2,3}$, Fe $L_{2,3}$, Ba $M_{4,5}$ and Pr $M_{4,5}$ absorption edges. All signals drop from 75% to 25% within distances below 0.5 nm. Having in mind that some beam broadening effects are present due to dechanneling [26], this finding points to the presence of atomically sharp interfaces. Finally, we could not detect a clear c-axis lattice expansion (beyond the noise level, i.e. in the few per cent units) across the YBCO-BFO interface, which is consistent with the absence of any appreciable interfacial concentration of oxygen vacancies in the virgin state. This is confirmed by estimates of the Fe oxidation state from EELS data, in particular from the Fe $L_{2,3}$ intensity ratio, which is constant (within the technique sensitivity) across the entire BFO layer.

Fig. 3(a) shows a HAADF image of a BFO-Mn (30 nm)/YBCO(5 u.c.)/PBCO(4 u.c.)/STO heterostructure. Given the Z-contrast nature of the image, the brightness of each atomic column is roughly proportional the square of its atomic number. The image shows the whole stacking with the BFO film on the top, deposited on a bilayer composed of five unit cells of YBCO and four unit cells of PBCO, grown on the STO substrate. We have carried out a quantitative analysis of such HAADF images, in order to directly measure the displacements of Fe columns relative to the center of the Bi sublattice in the as-grown (virgin) state [21,27]. Such displacements measured along the [001] and [100] directions, namely $d_z$ and $d_x$, are plotted in Fig. 3(b) and (c) respectively. In the color maps, each unit cell is represented by a pixel. The color maps are accompanied by a graph in which we plot the Fe off-center displacement averaged over each pixel line, $<d_z>$ and $<d_x>$. The sign of $d_x$ shows a small variability, particularly close the interface, which indicates the presence of nano-domains with different projection of the polarization vector along the [100] direction. However, as displayed in Fig. 3 (b), $d_z$ is homegeneous and positive, that is, the Fe columns exhibit a net displacement that points away from the interface. This finding means that the polarization vector along [001] points from the ferrite into the cuprate [28], which is the most common finding in all the studied in this series. As shown earlier for ferroelectric/oxide interfaces [20], the origin of this preferential polarization direction can be explained by the presence of rigid interface dipoles which result from the highly polar structure of both ferrite [20] and cuprate [29,30], and from their different polarity stacking along the growth direction [001]. In particular, see Fig. 2, in the studied heterostructures a nominally polar $FeO_2^-$ plane faces a nonpolar $BaO^0$ plane at the interface. The basal planes of the (incomplete) interfacial ferrite/cuprate unit cells are respectively $BiO^+$ and $CuO^-$ (as recently shown [30], the Cu valence in the CuO chains is close to +1). Thus, considering that the formal nominal charge of each basal plane is shared with a neighboring unit cell, the polar stacking can be described by the sequence $BiO^{+0.5}/FeO_2^-$/$BaO/CuO_2^{-1}/Y^{3+}/CuO_2^{-1}/BaO/CuO^{-0.5}$. As a consequence, the polarity mismatch yields a net nominal excess charge of +0.5/cell for the cuprate (due to the missing $CuO^-$ chains plane) and of -0.5/cell for the ferrite (due to the missing $BiO^+$ plane) incomplete interface cells. Thus, a pinned interfacial (non-switchable) dipole builds up, which points from the ferrite into the cuprate. From this finding, we expect a preferred polarization direction within the bulk of the BFO layer [20] pointing towards the YBCO layer as indeed observed in Fig. 3 (b). This preferred polarization direction is consistent with the observation of similar transport properties in the virgin state and after the BFO has been poled towards the YBCO layer [Fig. 1 (a)].

Fig. 3(d) displays the Fe off-center displacement within the (010) plane $D = \sqrt{d_x^2 + d_z^2}$ (color map) and the laterally averaged $<D>$ profile (adjacent panel). The mean value of the displacement is $D = 36 \pm 7$ pm. From this, the projection of the polarization vector on the image plane can be readily estimated [28] as $P_{xz} = 2.5 \cdot D$ µC cm$^{-2}$ = 90 ±18 µC cm$^{-2}$, with the projections along the axes being $P_z$ =70 ±12 µC cm$^{-2}$ and $P_x$ =50 ±18 µC cm$^{-2}$. Assuming that $P_x=P_y$ we estimate the spontaneous polarization $P$ =100 ±18 µC.cm$^{-2}$, which is in good agreement with the value for bulk BFO [28,31]. Note however that $P_z > P_x$, in contrast to as in bulk BFO for which $P_z = P_x = P_y \sim 60$ µC.cm$^{-2}$. That is, in the studied sample the polarization vector is slightly rotated from the [111] towards the [001] direction. This is as expected [31] (both qualitatively and quantitatively) from the existence of compressive epitaxial strain on the BFO, which amounts ~1.5% given the lattice parameters of STO (3.90 Å) and bulk BFO (3.96 Å) and considering that the few layers of YBCO/PBCO grow fully strained.

Fig. 3 (b) shows that the magnitude of $d_z$ (and hence $P_z$, which yields the field effects on YBCO) is homogeneous within the noise level. Particularly, $<d_z>$ at the interface is not much lower than far from it. Overall, $<d_z>$ variations are at most 30%. Altogether, the facts that the native *P* estimated from STEM is as expected for bulk BFO and $d_z$ (and thus $P_z$) is not substantially depressed near the interface rule out poor native ferroelectric properties, which consequently cannot account for any major reduction of the field-effect doping efficiency.

**C. Piezo-response microscopy**

A study of the ferroelectric properties at the microscale was carried out by piezo-response microscopy (PFM). The piezo-response phase (either 0º or 180º) indicates the polarity of the net polarization under the CT-AFM tip: towards the YBCO (180º) or outwards from it (0º). Fig. 4(a) shows a piezo-response phase image of a BFO/YBCO/PBCO//STO heterostructure in which a pattern – two concentric squares with homogeneous, opposite piezo-response phase – has been "written" [5]. The writing is done by first applying a dc voltage $V_{DC}$ = +5V between the YBCO layer and the CT-AFM tip (grounded) while this is scanned over the large square (dark contrast). Subsequently, $V_{DC}$ = -5 V is applied while scanning over the small square (bright contrast). Then, the PFM image is taken at remanence. One can see in Fig. 4 (a) a multidomain structure in the virgin area (around the largest square). Note however that bright-contrast domains (180º phase) are predominant. This finding suggests a preferred ferroelectric polarization polarity in the virgin state. Fig. 4 (c) shows piezo-response phase *vs.* $V_{DC}$ loops measured with the tip positioned at a fixed spot. One can see that $|V_{DC}|$ < 2V is sufficient to switch the phase to 180º (corresponding to a net polarization pointing towards the YBCO) while $|V_{DC}|$ ~ 3V is required to switch the phase to 0º. This asymmetry also suggests a preferential polarity, namely with the polarization vector along the [001] direction pointing from the BFO into the YBCO. This is fully consistent with the STEM observations discussed above Furthermore, an unusual, strong asymmetry is observed also in the piezo-response amplitude. Fig. 4 (b) exhibits the piezo-response amplitude image counterpart of Fig. 4 (a). The amplitude is significantly lower in the area where the phase is 0º (dark contrast, net polarization pointing away from away from the YBCO) than in the area where the phase is 180º (bright contrast). The same asymmetry is observed in the PFM amplitude *vs.* $V_{DC}$ cycle, see Fig. 4 (d). This behavior is consistently observed in a series of samples with varying BFO and YBCO thickness. Since the piezo-response amplitude scales with the modulus of the ferroelectric polarization $P$ (averaged under the CT-AFM tip), the stark contrast observed in Figs. 3 (b) and (d) implies that the effective $P$ is much smaller when the BFO is poled outwards from the YBCO layer ($|P_\uparrow|$) than when it is poled towards the YBCO ($|P_\downarrow|$). In other words, the amplitude contrast suggests that full ferroelectric switching is not achieved upon poling, and consequently that the polarization variation $\Delta P = P_\uparrow - P_\downarrow$ must be significantly smaller than twice the virgin BFO polarization projected along the [001] direction, $2P_z \sim 140$ μC cm$^{-2}$. This fact must largely contribute to the reduced field-effect doping efficiency discussed in section A, which is the main conclusion of the PFM study.

With the evidence at hand, the low piezo-response amplitude (= small effective $|P_\uparrow|$) when the ferroelectric is poled outwards from the YBCO can be understood if we consider the existence of pinned ferroelectric domains in which the polarization has the preferred polarity as suggested earlier [32,33]. In particular, we can propose the scenario sketched in Fig. 4 (e), in which the ferroelectric polarization under the AFM tip is homogenous after being poled towards the YBCO layer, but shows residual domains with the preferred polarity after being poled in the opposite direction. There may be alternative scenarios to explain the small effective $|P_\uparrow|$. For example, a depression of the spontaneous polarization due to charge transfer effects [34] or to the migration of oxygen vacancies into BFO upon ferroelectric poling. To discriminate between the possible explanations, further experiments would be required, for instance STEM studies (as those discussed above) as a function of the ferroelectric poling history.

**IV. THEORETICAL MODEL AND COMPARISON TO EXPERIMENT**

Fig. 5 (a) displays a toy model that allows for a quantitative understanding of the transport experiments, and includes the learnings from the STEM and PFM studies discussed above.

The YBCO layer is divided in two layers of thickness $t_1$ and $t_2$, with the total thickness $t_{YBCO} = t_1 + t_2$. The critical temperature T$_{C2}$, the carrier density $n_2$ and mobility $\mu_2$ are uniform and independent of the ferroelectric polarization far enough from the interface (layer "2"). Contrarily, close to the interface (layer "1") $n_1$, T$_{C1}$ and $\mu_1$ depend on the ferroelectric polarization direction. Note that, although we expect an exponential decay of the accumulation of screening charges [35,36] and consequently a gradient of the electronic properties across layer "1", for simplicity we consider uniform $n_1$, T$_{C1}$ and $\mu_1$ for each polarization state ($P_\uparrow$ and $P_\downarrow$). This approximation does not impact the conclusions drawn below. Note finally that, in this model, $t_1$ is virtually equivalent to the Thomas Fermi screening length $\lambda_{TF}$.

Based on that model, and using a standard approach for multilayer systems [37], the conductivity $\sigma'$ and mobility $\mu'$ expected in a measurement of the YBCO film are given by

$$\sigma' t_{YBCO} = \sigma_1 t_1 + \sigma_2 t_2 \tag{Eq. 1}$$

$$\mu' \sigma' t_{YBCO} = \mu_1 \sigma_1 t_1 + \mu_2 \sigma_2 t_2 \tag{Eq. 2}$$

Using the relationships for the conductivities $\sigma' = \mu' n' e$, $\sigma_i = \mu_i n_i e$ and the sheet carrier density $n_{2D_i} = t_i n_i$ (with $i = 1,2$), the Hall coefficient $R_H = 1/en'$ expected in an measurement obeys the relationship:

$$t_{YBCO}/eR_H = \frac{\left(\frac{\mu_1}{\mu_2} n_{2D_1} + n_{2D_2}\right)^2}{\left(\frac{\mu_1}{\mu_2}\right)^2 n_{2D_1} + n_{2D_2}} \tag{Eq. 3}$$

From this equation, one readily realizes the measured $t_{YBCO}/eR_H$ equals the sheet carrier density in the YBCO layer $n_{2D} = n_{2D_1} + n_{2D_2}$ only if $\mu_1 = \mu_2$. Otherwise, $t_{YBCO}/eR_H \neq n_{2D}$ and, consequently, the Hall coefficient variation observed upon ferroelectric switching is not a direct measurement of the sheet carrier density variation, that is, $t_{YBCO} \cdot |\Delta(1/R_H)| \neq e\Delta n_{2D} = \Delta P$. Considering this and the fact that $\Delta P$ must be very similar for all the samples in the series, the steady decrease of $t_{YBCO} \cdot |\Delta(1/R_H)|$ with increasing $t_{YBCO}$ [Fig. 1 (c)] indicates that the carrier mobility varies across the YBCO layer ($\mu_1 \neq \mu_2$) at least for one of the states (accumulated/depleted). This scenario is further confirmed by numerical simulations based on Eq. 3, which allow us to quantitatively reproduce the experimental results.

In order to simulate the experimental behavior via model calculations, we proceeded as follows. Since the native carrier concentration $n_2$ is not the same in all the samples (see in Fig. 1 (b) that $1/eR_H$ is modulated around ~$10^{21}$ cm$^{-3}$ for some samples and around ~$2 \cdot 10^{21}$ cm$^{-3}$ for others) we made calculations in the range $10^{21}$ cm$^{-3}$<$n_2$<$2 \cdot 10^{21}$ cm$^{-3}$. We accordingly adjusted the thickness of region "1" between 3 u. c.> $t_1$ >2 u. c. Note that $t_1$ is not a fitting parameter. We fixed $t_1$ for each $n_2$ by considering that the screening length is $\lambda_{TF}$ ~ 1 nm (~1 u.c.) for optimally doped YBCO ($n$~ $6 \cdot 10^{21}$ cm$^{-3}$) [35,36] and that this length is inversely proportional to the square-root of the carrier density [35,36]. Then we searched $n_1$ and $\mu_1/\mu_2$ (in the accumulated and depleted states) that allowed to mimic the decay of $t_{YBCO} \cdot |\Delta(1/R_H)|$ seen in Fig. 1 (c), by imposing two consistency conditions. First, the same $e\Delta n_{2D_1} = e\Delta n_{2D} = \Delta P$ is used for all $n_2$, which means that the ferroelectric properties do not change from sample to sample, as indicated by their similar PFM and STEM properties. Second, we impose $|n_1 - n_2|_{accumulated} < |n_1 - n_2|_{depleted}$, which accounts for the assymetric behavior $|P_\uparrow| < |P_\downarrow|$ evidenced by PFM. The simulation results are represented by lines in Fig. 1 (c). The continous horizontal line corresponds to the case in which $\mu_1/\mu_2 = 1$ both in the depleted and accumulated states. Here $t_{YBCO} \cdot |\Delta(1/R_H)|$ is constant, independent of $n_2$, and equals the chosen $e\Delta n_{2D} = 32$ µC.cm$^{-2}$. The dashed and dash-dotted lines that envelope the experimental data respectively correspond to calculations for $n_2 = 1.10^{21}$ cm$^{-2}$ and $n_2 = 1.9 \; 10^{21}$ cm$^{-2}$ with the fitting parameters $e\Delta n_{2D} = 32$ µC.cm$^{-2}$, $\mu_1/\mu_2(accumulated) = 0.36$ and $\mu_1/\mu_2(depleted) = 1$, which yield the best match between model calculations and experimental data. There is some degree of uncertainty in those parameters. For example, similar results are obtained with $0.7 < \mu_1/\mu_2(depleted) \leq 1$, or with up to 20% higher $e\Delta n_{2D}$ at the expense of using lower $\mu_1/\mu_2(accumulated)$. However, a good matching can be obtained only if i) $e\Delta n_{2D} < 50$ μC cm$^{-2}$ and ii) $\mu_1/\mu_2 (accumulated) < \mu_1/\mu_2 (depleted) \leq 1$ (this is illustrated in the Supplemental Material by a series of simulations with varying fitting parameters). In other words, the modelling i) confirms that $\Delta n_{2D} \cdot e \ll 2P_z \sim 140$ μC cm$^{-2}$ and ii) shows that the YBCO carrier mobility close to the interface (layer "1") is significantly reduced and varies upon poling the ferroelectric.

Further comparison between model calculations and experiments is done in Figs. 5 (b)-(c). The data points come from a series of samples grown in a single run and having a comparable native carrier density $n_2 \sim 10^{21}$ cm$^{-3}$. Fig. 5 (b) shows $1/eR_H$ vs. $t_{YBCO}$ measurements in the depleted/accumulated states (respectively circles/squares). One can see that in the accumulated state (squares) $1/eR_H$ drops rapidly with increasing $t_{YBCO}$ and tends to merge $1/eR_H$ in the depleted state (circles). Calculations (dashed and continuous lines) are made with the same parameters as in Fig. 1 (c) ($n_1$ in the depleted/accumulated states and $n_2$ are respectively indicated by blue, red and black horizontal dotted lines). The continuous curves correspond to calculations with uniform mobility across the YBCO layer ($\mu_1/\mu_2 = 1$). The dashed curves correspond to the case in which $\mu_1$ is reduced upon carrier accumulation ($\mu_1/\mu_2 = 0.36$). This simulation case closely reproduces the experimental trend, particularly the marked decay of $1/eR_H$ in the accumulated state (squares) which is not accounted for by the simulation case $\mu_1 = \mu_2$ (continuous lines). The stark contrast between the behavior of the dashed and continuous curves illustrates once again the sensitivity of the Hall coefficient to carrier mobility variations across the YBCO layer. That sensitivity is an effect of the electrical current redistribution between layers "1" and "2" [37]. Note finally that the scattering of the experimental data (squares/circles) around the theoretical curves (dashed) reflects the variability of $n_2$ from sample to sample, which seemingly amounts 10% - 15%. Consistently, that scattering disappears in the representation of in Fig. 5 (c), which shows the variation between the accumulated and doped state for the same data and calculations displayed in (a).

Within the model in Fig. 5(a), the critical temperature in layer "1" $T_{C1}$ will depend on the polarization state (accumulated/depleted) since, as discussed above [see $T_c(n)$ in Fig. 1 (b)], in YBCO the critical temperature depends on the carrier concentration [22–25]. In layer "2" the carrier density (and thus $T_{C2}$) does not change upon ferroelectric poling. Note that the expected $T_C$ (as measured with the electrical current injected parallel to the interface) equals the maximum between $T_{C1}$ and $T_{C2}$. Thus, in the doped state, the measured $T_C$ equals $T_{C1}$. Contrarily, in the depleted state $T_{C1}$ is depressed, and thus the measured $T_C$ equals $T_{C2}$. This allows for a qualitative explanation of the experimental results in Fig. 5 (d). One can see that $T_C$ in the accumulated state does not depend on $t_{YBCO}$. This is because $t_1$, $n_1$ and thus $T_{C1}$ are independent on $t_{YBCO}$. In the depleted state $T_C = T_{C2}$ strongly decreases with decreasing $t_{YBCO}$. This is understood if one considers that $t_2 = t_{YBCO} - t_1$ decreases with decreasing $t_{YBCO}$. Thus one expects that $T_{C2}$ decreases with decreasing $t_{YBCO}$, as observed, because the critical temperature of ultrathin YBCO films decreases with decreasing thickness (one can think of $t_2$ as an "isolated" ultrathin film since the critical temperature in $t_1$ is strongly depressed/suppressed in the depleted state).

## V. CONCLUSIONS

We have unveiled and characterized various factors that limit the ferroelectric field-effect doping efficiency in the $BiFeO_3/YBa_2Cu_3O_{7-\delta}$ heterostructures. First, we have shown that full ferroelectric switching is not achieved after electrical poling, which clearly contributes to the carrier density modulation being below the maximum expected value $2P$ (with $P$ the bulk ferroelectric polarization normal to the interface). The incomplete ferroelectric switching is likely connected with the existence of a preferential polarization direction, as denoted by both PFM and STEM experiments. The preferred direction is set by the atomic terminations at the interface, which result in a fixed interface dipole pointing from the ferrite into the cuprate. In this scenario, the weaker polarization measured when the ferroelectric is poled opposite to the preferential direction could be explained by the presence of residual domains in which the polarization remains pinned. These observations point to interface engineering as a key route to enhance ferroelectric field effects.

Second, and more importantly, our experiments show that the field-effect induced carrier accumulation is accompanied by a substantial change of the carrier mobility in the oxide channel. This finding is relevant in two ways:

i) On the one hand, it shows that quantitative studies of field-effect doping through Hall experiments are not straightforward, since changes in the Hall coefficient upon ferroelectric switching do not only reflect the carrier density modulation, but also the changes in the carrier mobility. Especially when the channel is thicker than the Thomas-Fermi screening length, neglecting changes of the mobility leads to misestimating the carrier density modulation. In the present experiments, if the mobility decrease is neglected the carrier density modulation is underestimated, to an extent that is greater the thicker the YBCO.

ii) On the other hand, the mobility variations produced by ferroelectric poling suggest that the doping effects are accompanied by structural changes in the YBCO channel. A consistent scenario would be that, upon ferroelectric poling, charged defects such as oxygen vacancies are induced [16,21]. Structural distortions controlled by the polarization state could also contribute to the mobility variations, as recently shown [19]. Further experiments, for example STEM studies as a function of the polarization state, would be required to ascertain the physical origin of the mobility variations.

Finally, we stress that the factors discussed above may play a relevant role in other ferroelectric/complex-oxide interfaces, because the key ingredients (such as the presence of polar interfaces, oxygen vacancies, or the oxide channel being thicker that the Thomas-Fermi screening length) are ubiquitous in these systems and their studies.

**Acknowledgements**

Work supported by the ERC grant Nº 64710, French ANR grant ANR-15-CE24-0008-01 and European COST action 16218 "nanocohybri". Work at UCM supported by Spanish MINECO-FEDER MAT2015-66888-C3-3-R and ERC PoC2016 POLAR-EM. Electron microscopy observations at ORNL were supported by the U.S. Department of Energy, Basic Energy Sciences, Materials Sciences and Engineering Division. J. S. thanks INP-CNRS and "Scholarship program Alembert" funded by the IDEX Paris-Saclay ANR-11-IDEX-0003-02 for support during his stay at the Unité Mixte de Physique CNRS/Thales. J.E.V. thanks S. Fusil and V. Garcia for discussions and suggestions.

**Fig. 1: (a)** Resistance versus temperature of a BFO-Mn (30 nm)/YBCO(4 u.c.)/PBCO(2 u.c.)/STO// heterostructure, measured with an injected current J=230 A·cm$^{-2}$ in the virgin state (green curve), after poling the ferroelectric towards the YBCO (depleted state, blue curve) and away from the YBCO (accumulated state, red curve). The inset shows a sketch of the lithographed multi-probe bridge used for electrical measurements. In the central area (between the four voltage probes $V_1$-$V_4$), the remnant ferroelectric polarization is set by applying $|V_{DC}|$ (typically 5 V) between the scanning AFM tip and the YBCO **(b)** Critical temperature $T_C$ versus the $1/eR_H$ for a series of samples with the same BFO thickness (30 nm) and variable YBCO thickness (symbols' legend in unit cells). For a given sample (symbol), red/blue corresponds to the accumulated/depleted state, and green to the virgin state. The dashed curve is the $T_C$ versus charge carrier density $n$ for bulk YBCO. The experimental $T_C$ is defined as the temperature at which the resistance drops to 90% of that at the onset of transition. **(c)** YBCO thickness $t_{YBCO}$ times the variation $\Delta(1/R_H)$ between depleted and doped states, plotted as a function of $t_{YBCO}$. The shaded area is a guide to the eye. The dashed lines correspond to simulations with the model of Fig. 5.

**Fig. 2 (a)** High resolution HAADF image of a BFO-Mn (30 nm)/YBCO(3 u.c.)/PBCO(4 u.c.)/STO sample (top), along with a sketch of the proposed atomic plane stacking sequence at the BFO/YBCO interface (bottom). Y, Cu, Ba, Fe and Bi atomic columns are shown in dark blue, blue, green, red and light brown, respectively. **(b)** EELS atomic resolution maps across the stacking, obtained from MLLS fits of raw EEL spectrum images. From top to bottom : Ti $L_{2,3}$, Fe $L_{2,3}$, Ba $M_{4,5}$ and Pr $M_{4,5}$ maps, along with an overlay of the Ti (yellow), Fe (red), Ba (green) and Pr (blue) signals. Some spatial drift is observed. The panel at the bottom exhibits the laterally averaged normalized signals along the direction marked with a blue arrow in **(a)** corresponding to the four chemical signals, same color scale. Vertical dashed lines mark the position of the interfaces, while horizontal maps highlight the 25% and 75% levels.

**Fig. 3 (a)** STEM-HAADF cross-sectional image of an as-grown BFO-Mn (30 nm)/YBCO(5 u.c.)/PBCO(4 u.c.)/STO// heterostructure. **(b)**, **(c)** and **(d)** panels exhibit false color maps of the off-center displacements of Fe columns relative to the center of the Bi sublattice, measured for every unit cell (represented by a pixel) along the [001], [100] directions and in the (010) plane, respectively. A positive $d_z$ means that the Fe columns exhibit a net displacement that points away from the interface. The maps correspond to the BFO area framed with blue dashed lines in **(a)**. Each color map is accompanied by a profile of the Fe displacement averaged laterally over every whole pixel line. All the scales are in Angströms.

**Fig. 4 (a)** PFM phase and **(b)** amplitude image of the surface of a BFO-Mn (30 nm) /YBCO (4 u.c.)/PBCO (4 u.c.)/STO// heterostructure. The image size is 6 μm × 6 μm. The background corresponds to the virgin state. A square with opposite PFM phase contrast (dark) has been "written" by applying a dc voltage $V_{DC}$ = 5 V between the YBCO layer and the scanning tip (ground). The phase contrast is reversed again in a concentric square (bright) by applying $V_{DC}$ = -5 V **(c)** Local piezo-response phase and amplitude **(d)** as a function of $V_{DC}$ (arrows and labels indicate the $V_{DC}$ cycling sequence). Black and blue symbols correspond to two subsequent cycles. Similar loops were observed upon repeated cycling, and also for different positions of the tip. **(e)** Cartoon of the proposed remnant ferroelectric domain structure after application of a large negative/positive voltage. This allows for an understanding the piezo-response amplitude contrast behavior observed in (b) and (d).

**Fig. 5: (a)** Cartoon of the field-effect model at the ferroelectric/superconductor interface. When the ferroelectric polarization ***P*** points away from the YBCO (accumulated state), the charge (hole) accumulation leads to a carrier density increase over some length scale, locally rising $T_C$. The native carrier density (and $T_C$) are recovered further from the interface. This is modelled by virtually separating the superconductor into two regions of thickness $t_1$ and $t_2$, the former

having an average carrier density $n_1$ higher than the latter $n_2$. The $T_C$ of the superconductor, as probed by electron transport measurements in the direction parallel to the interface, corresponds to that of the doped region $T_{C1}$. On the contrary, upon polarization reversal (depleted state), the measured $T_C$ corresponds to that of the region far from the interface $T_{C2}$ which is the highest because the depleted $n_1 \ll n_2$. **(b)** Inverse of the Hall coefficient as a function of the total thickness of the superconductor $t_{YBCO}=t_1+t_2$, for the doped (red) and depleted (blue) states. The symbols (squares/circles) are experimental data. The solid lines correspond to model calculations for the case in which the carrier mobility $\mu$ is uniform across layers 1 and 2. The dashed curves correspond to calculations for the case in which the mobility close to the interface $\mu_1$ varies upon ferroelectric switching. The horizontal lines mark $n_1$ (red/blue respectively for the doped/depleted state) and $n_2$ (black, independent of the polarization state). **(c)** Variation between the depleted and accumulated states for the same data and calculations as in (b). **(d)** Critical temperature measured for the same samples as in (b). The lines are a guide to the eye.

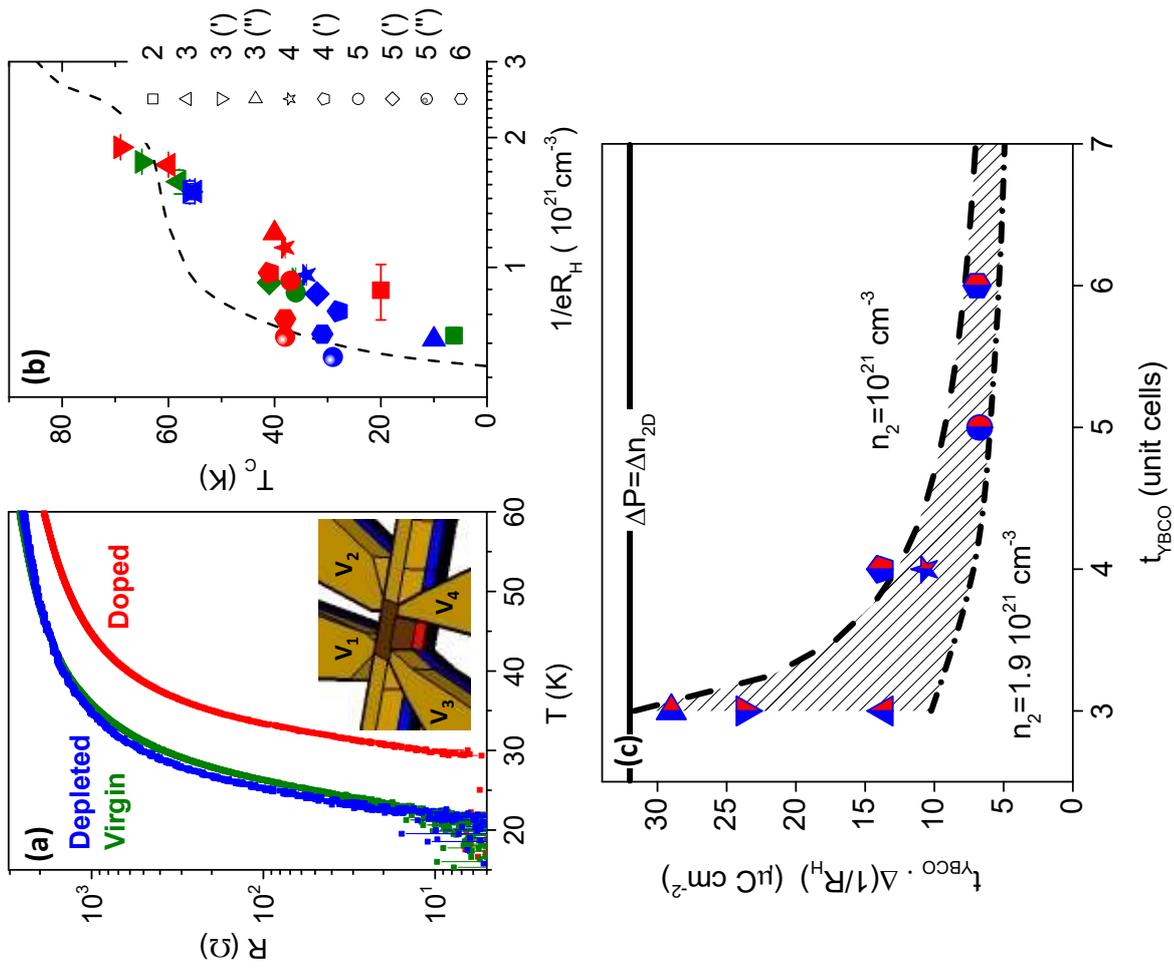

FIGURE 1

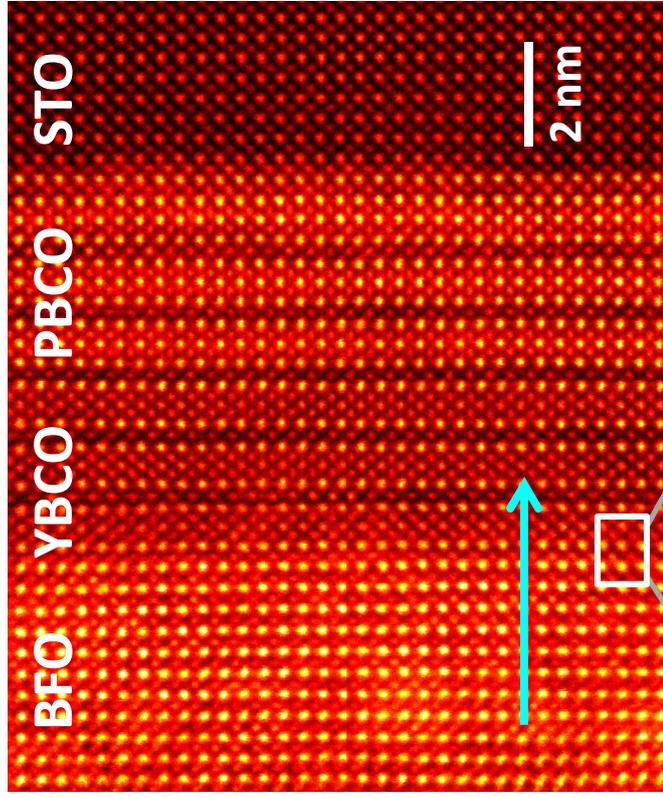
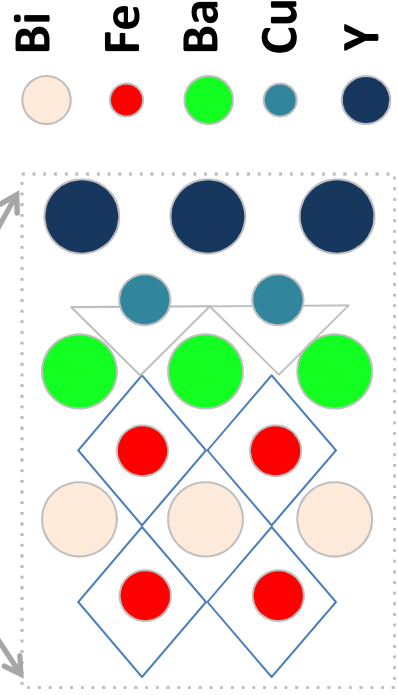
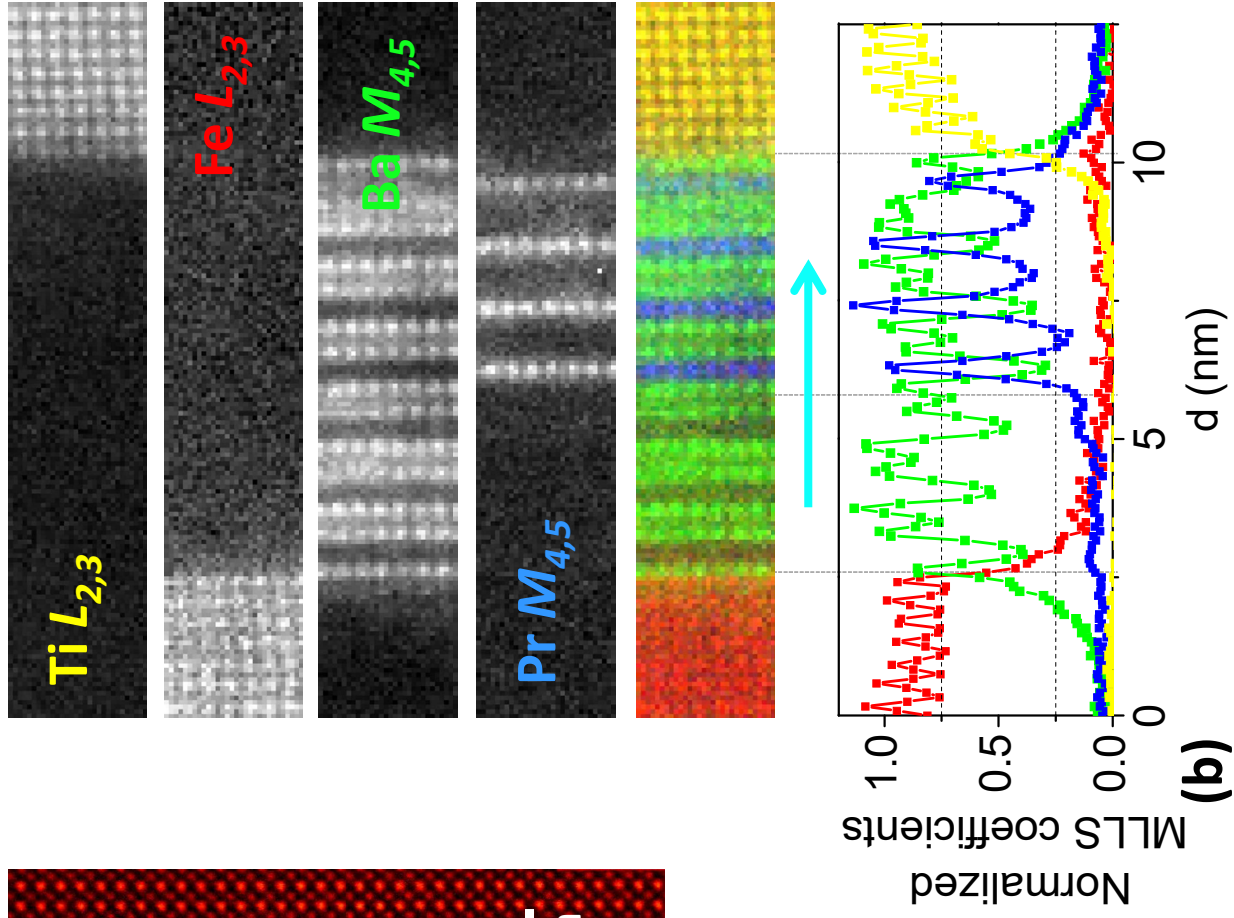

FIGURE 2

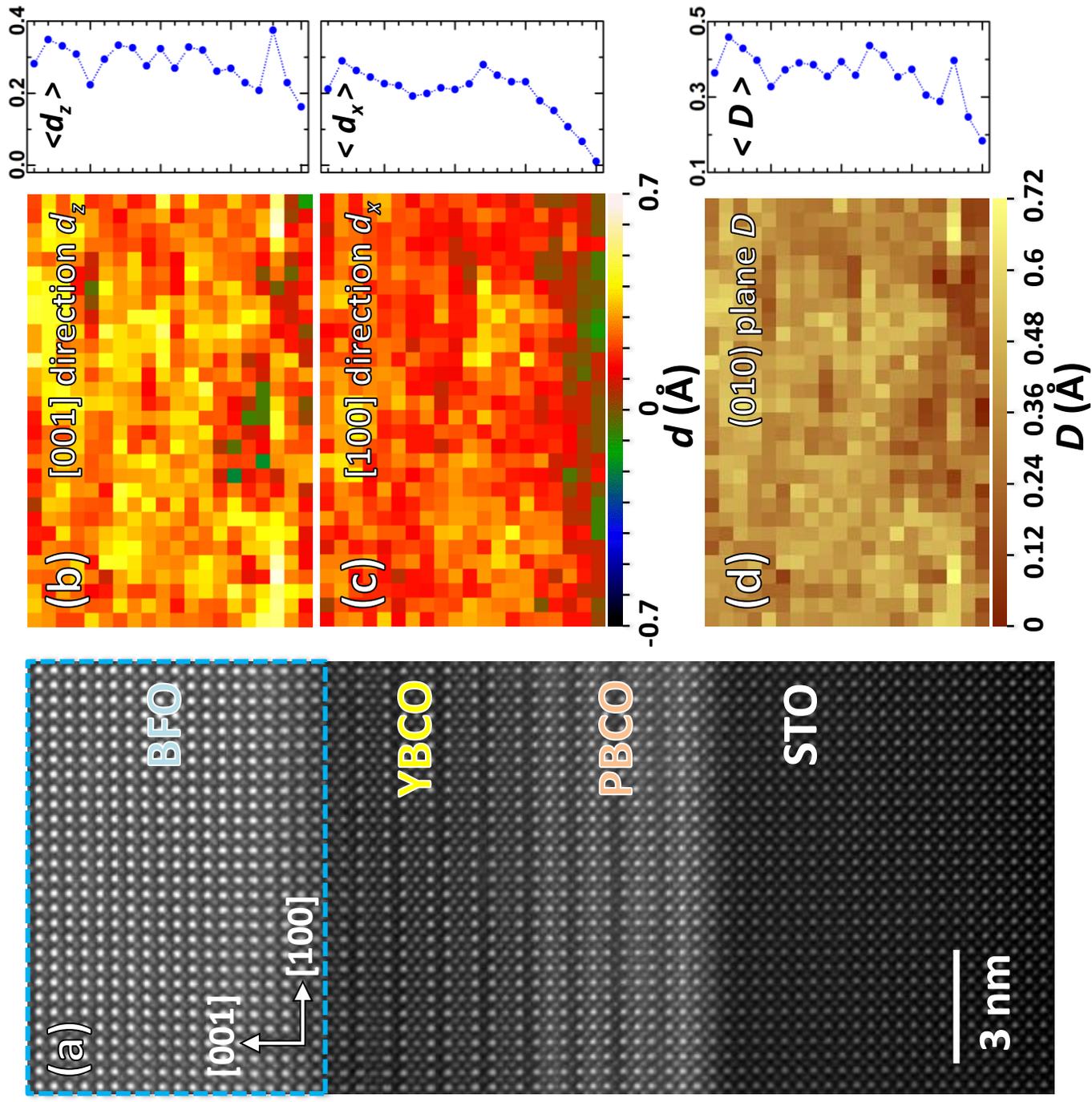

FIGURE 3

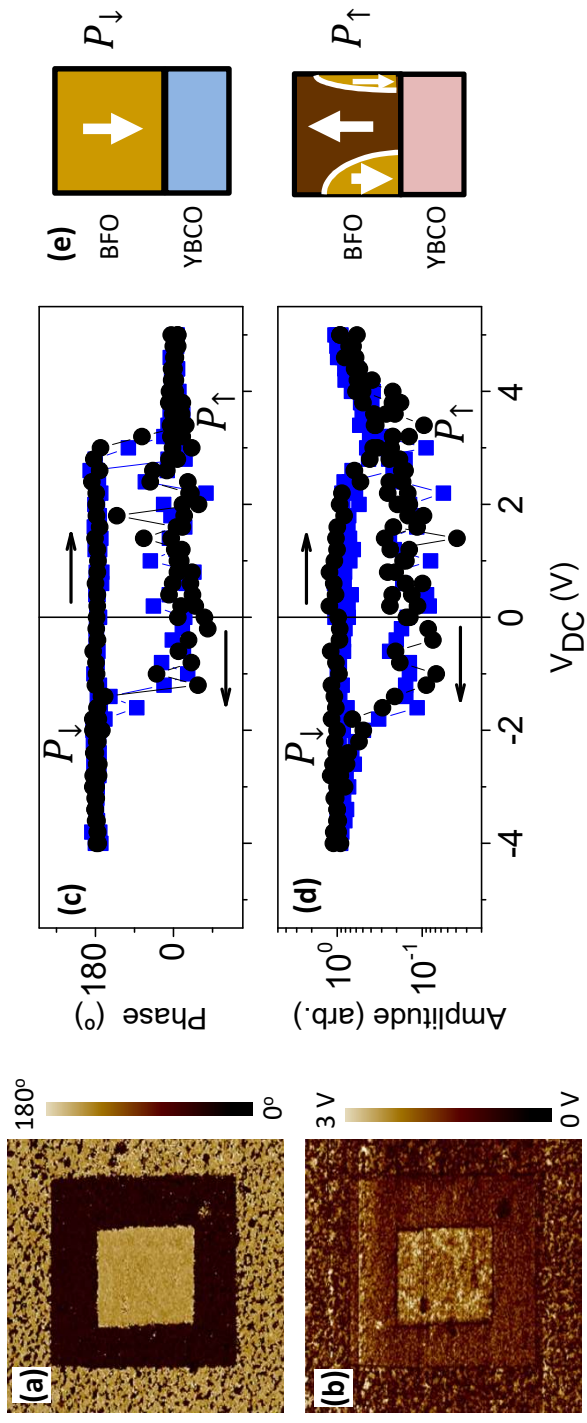

FIGURE 4

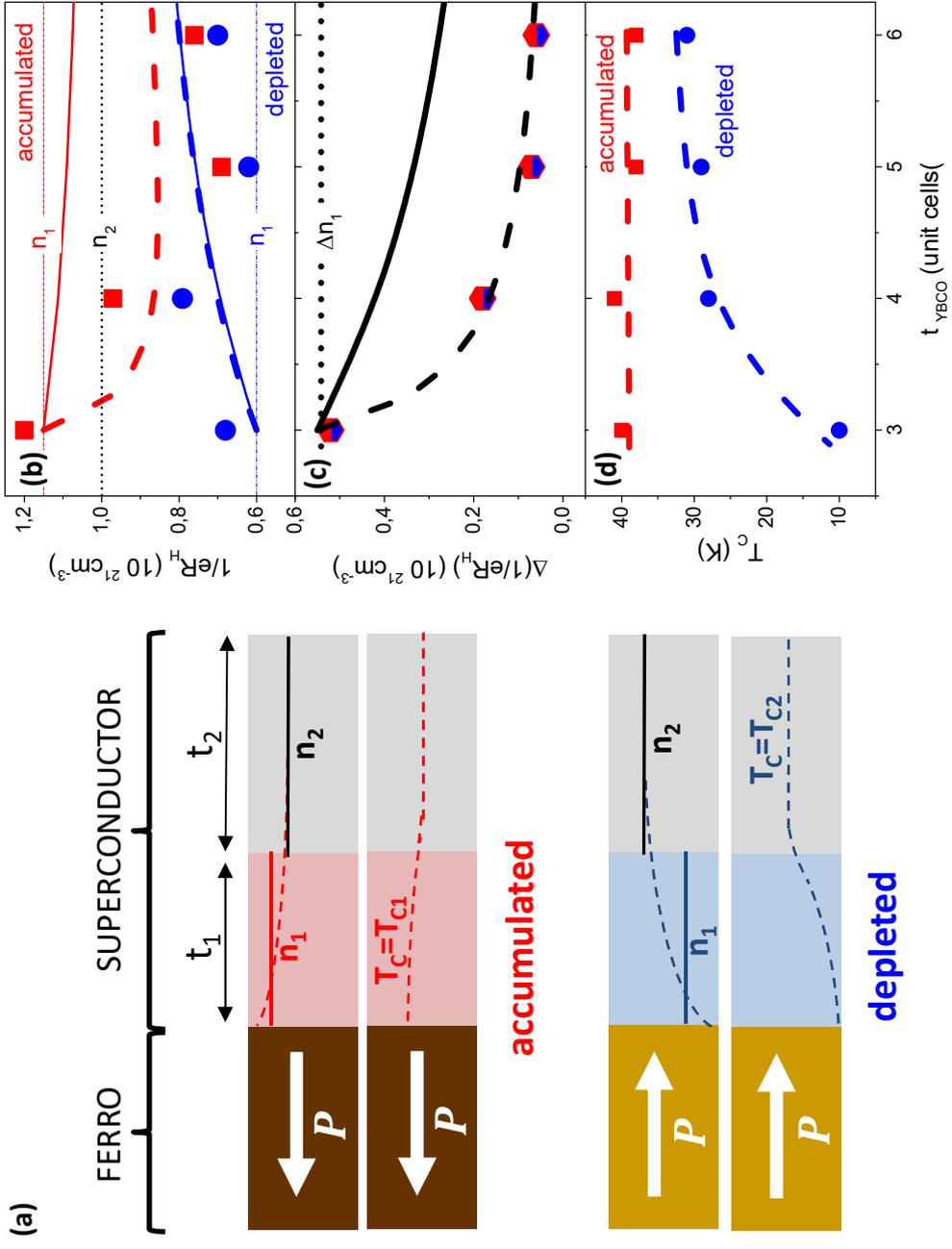

FIGURE 5

# Supplemental Material for

# Factors limiting ferroelectric field-effect doping in complex-oxide heterostructures

L. Bégon-Lours[1,♦], V. Rouco[1], Qiao Qiao[2], A. Sander[1], M. A. Roldán[3], R. Bernard[1,♣], J. Trastoy[1], A. Crassous[1], E. Jacquet[1], K. Bouzehouane[1], M. Bibes[1], J. Santamaría[1,3], A. Barthélémy[1], M. Varela[3] and Javier E. Villegas[1,*]

[1]Unité Mixte de Physique, CNRS, Thales, Univ. Paris-Sud, Université Paris Saclay, 91767 Palaiseau, France
[2]Oak Ridge National Laboratories, Oak Ridge TN 37840, USA
[3]GFMC, Dpto. Física de Materiales, Universidad Complutense de Madrid, 28040, Spain

* corresponding author: javier.villegas@cnrs-thales.fr
♦ present address: MESA+ Institute, Inorganic Materials Science, Enschede, The Netherlands
♣ present address: Fonctions Optiques pour les Technologies de l'Information, UMR 6082, CNRS, INSA de Rennes, Rennes, France

# 1. Sensibility of the model calculations to variations of the ratio between mobilities $\mu_1/\mu_2$.

In order to illustrate the sensibility of the model calculations to the parameter $\mu_1/\mu_2$, in the following pages we will show simulations for various combinations of $\mu_1/\mu_2(accumulated)$ and $\mu_1/\mu_2(depleted)$, and compare them to the best fit which is obtained for $\mu_1/\mu_2(accumulated) = 0.36$ and $\mu_1/\mu_2(depleted) = 1$.

Figure S1 shows that a good agreement between calculations and experimental data is obtained when i) in the depleted state, the carrier mobility $\mu_1$ nearby the interface [layer "1" in Fig. 5 (a)] is equal or lower than $\mu_2$ (far from the interface) and ii) $\mu_1$ is further decreased upon switching into the accumulated state. That is, that best agreement is obtained for $\mu_1/\mu_2 \leq 1$ and $\mu_1/\mu_2(accumulated) < \mu_1/\mu_2(depleted)$.

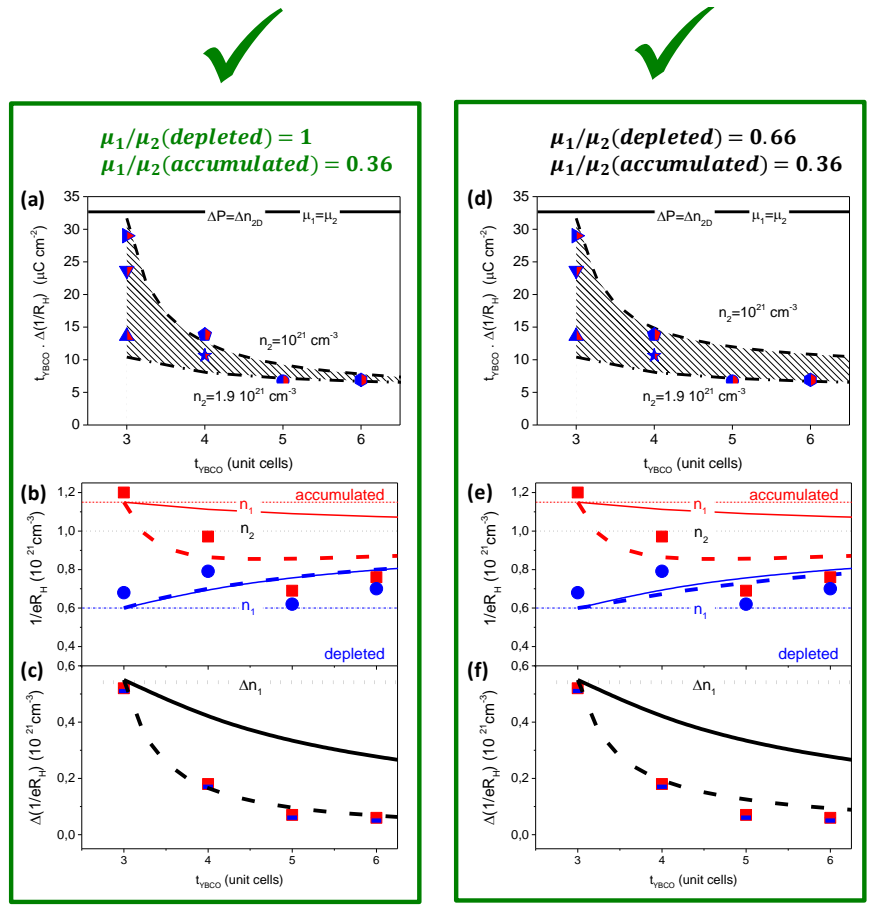

**Figure S1:** **(a)** YBCO thickness $t_{YBCO}$ times the variation $\Delta(1/R_H)$ between depleted and doped states, plotted as a function of $t_{YBCO}$. The shaded area is a guide to the eye. The dashed lines correspond to simulations with the model of Fig. 5 and the optimum fitting parameters, which include the set of $\mu_1/\mu_2(accumulated)$ and $\mu_1/\mu_2(depleted)$ indicated in top label. **(b)** For a series of samples with similar $n_2 \sim 10^{21}$ cm$^{-3}$, inverse of the Hall coefficient as a function of the total YBCO thickness $t_{YBCO}$, for the doped (red) and depleted (blue) states. The symbols (squares/circles) are experimental data. The solid lines correspond to model calculations for the case in which the carrier mobility $\mu$ is uniform across layers 1 and 2. The dashed curves correspond to calculations for the case in which the mobility close to the interface $\mu_1$ varies upon ferroelectric switching. The horizontal lines mark $n_1$ (red/blue respectively for the doped/depleted state) and $n_2$ (black, independent of the polarization state). **(c)** Variation between the depleted and accumulated states for the same data and calculations as in (b). **(d)-(e)-(f)** show the same as in (a)-(b)-(c) but with different set a of $\mu_1/\mu_2$ (see the top label), all the other parameters being identical.

Figs. S2 (d)-(i) allow comparison between experimental data and model calculations for other combinations of $\mu_1/\mu_2$. In particular for cases in which the mobility near the interface is equal or lower than far from it, that is $\mu_1/\mu_2 \leq 1$. The optimum fitting shown above in Fig. S1 is shown again [(a)-(b)-(c)] to compare with two representative cases. First, one in which the mobility increases when poling into the accumulated state [(d)-(e)-(f), see used $\mu_1/\mu_2$ in the top label]. Second, an example in which the mobility does not change upon poling [(h)-(i)-(j), see label]. One can see that in the both cases there is no agreement between calculations and experimental data. In particular, in (d) the calculations display a dependence with the YBCO thickness which is just opposite to the experimental trend. In (g), the calculated decay with YBCO thickness is much weaker than in the experiments. The disagreement is also clear in (e)-(f) and (h)-(i).

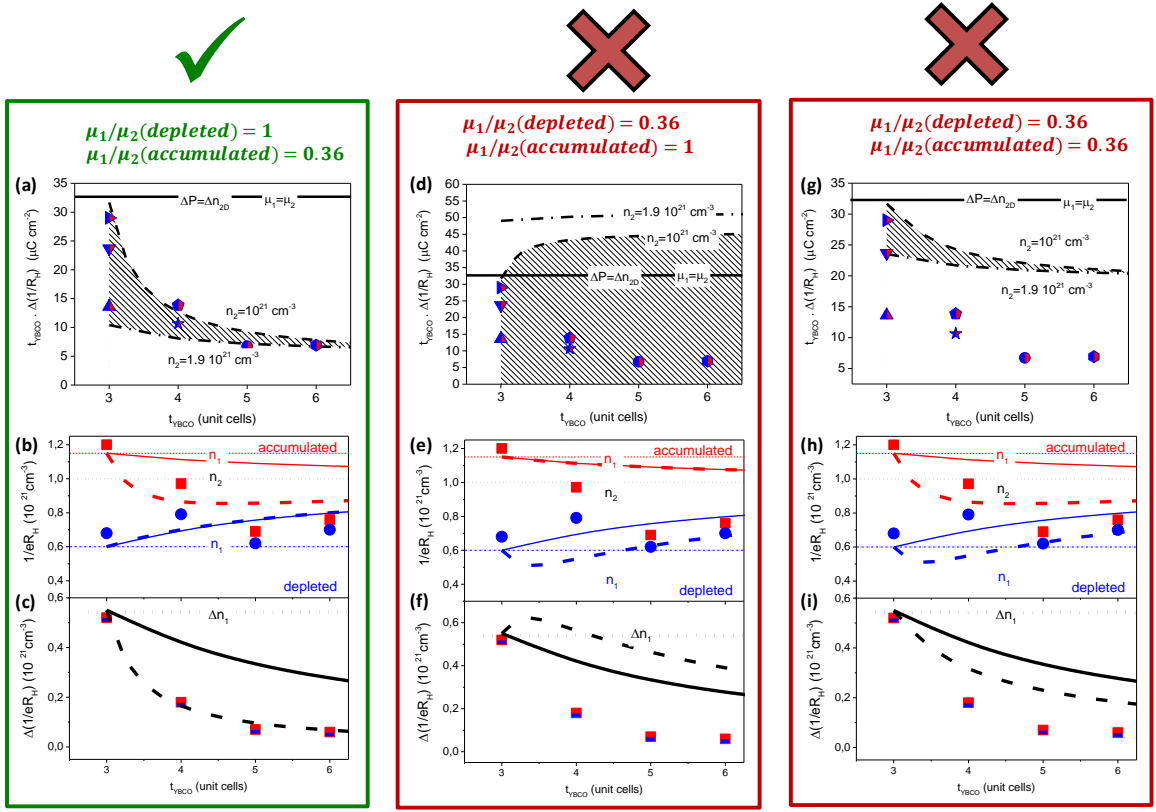

**Figure S2:** Same as in Fig S1 for **(a)-(b)-(c)** optimum set of parameters; **(d)-(e)-(f)** the mobility near the interface increasing upon doping (see the top label, all the other parameters being identical); **(g)-(h)-(i)** the mobility does not change upon poling (see the top label, all the other parameters being identical).

Fig. S3 allows comparison of the optimum fitting [(a)-(b)-)-(c)] with calculations for which the mobilities are in the opposite limit, that is $\mu_1/\mu_2 \geq 1$. One can see that in this limit no good agreement between calculations and experimental data can be obtained regardless of whether we consider that the mobility near the interface increases [(d)-(e)-(f)] or decreases [(g)-(h)-(i)] in the accumulated state. In particular, in (d) the decay with increasing YBCO thickness occurs at a much slower, steady rate in the calculations than in the experimental data. This is because the fast drop observed in (e) for the experimental data (in the accumulated state, red squares) is not reproduced by the calculations. In (g) we observe that the calculations show a trend which is just opposite to the experimental one.

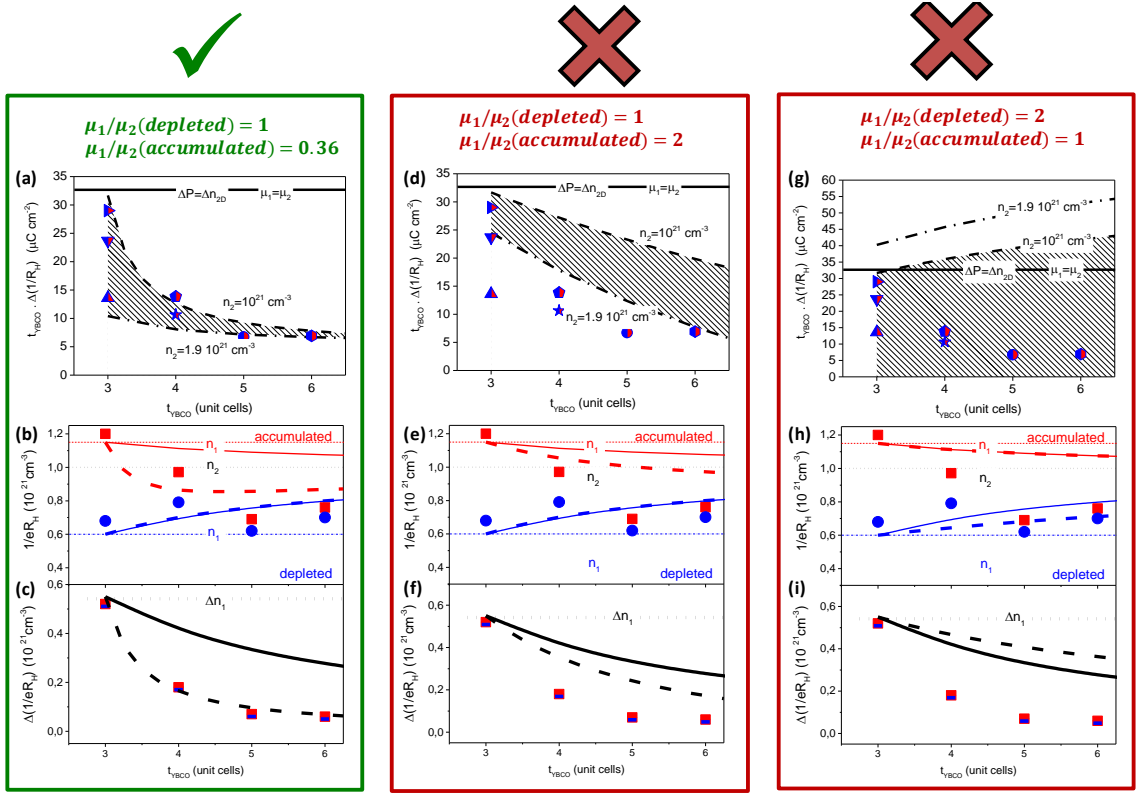

**Figure S3:** Same as in Fig S1 for **(a)-(b)-(c)** optimum set of parameters **(d)-(e)-(f)** $\mu_1/\mu_2 \geq 1$ and the mobility near the interface increasing upon doping (see the top label, all the other parameters being identical). **(g)-(h)-(i)** the mobility decreases in the accumulated state (see the top label, all the other parameters being identical).